\documentclass[journal]{IEEEtran}

\ifCLASSINFOpdf
\else
   \usepackage[dvips]{graphicx}
\fi
\usepackage{url}

\usepackage{graphicx}
\usepackage{amsmath}
\usepackage{algorithmic,amsfonts}
\usepackage{algorithm}
\usepackage{cite,float}

\usepackage{xcolor}
\usepackage{float}

\ifCLASSOPTIONcompsoc
\usepackage[caption=false,font=normalsize,labelfont=sf,textfont=sf]{subfig}
\else
\usepackage[caption=false,font=footnotesize]{subfig}
\fi

\begin{document}
	
\title{Point Cloud Resampling via Hypergraph Signal Processing}
\author{Qinwen~Deng, \IEEEmembership{Student Member, IEEE},~Songyang~Zhang, \IEEEmembership{Student Member, IEEE}, and~Zhi~Ding, \IEEEmembership{Fellow, IEEE}
\thanks{Qinwen Deng, Songyang Zhang and Zhi Ding are with the Department of Electrical and Computer Engineering, University of California, Davis, CA, USA. (e-mails: mrdeng@ucdavis.edu; sydzhang@ucdavis.edu).}}

\maketitle

\begin{abstract}
	Three-dimensional (3D) point clouds are
	important data representations in 
visualization applications. 
	The rapidly growing
	utility and popularity of point cloud processing strongly motivate a plethora of
	research activities on large-scale point cloud processing and feature extraction. In this work, we investigate point cloud resampling based on hypergraph signal processing (HGSP).
	We develop a novel method to extract sharp object features and reduce the data size of point cloud representation. 
By directly estimating hypergraph spectrum based on hypergraph stationary processing, we design a spectral 
kernel-based filter to capture high-dimensional interactions among 
point signal nodes and to better preserve object
surface outlines. Experimental results validate 
the effectiveness of hypergraph in representing point clouds, 
and demonstrate the robustness of the proposed algorithm
under noise. 
\end{abstract}


\section{Introduction}

\IEEEPARstart{T}{he} widespread deployment of networked sensors and cameras is a critical element of IoT. 
In visualization applications,
3D point clouds provide efficient data
representation of external surfaces for objects 
and their surroundings. 
Point clouds have found wide applications in recent years, including surface reconstruction \cite{d1}, 
rendering \cite{d2} and feature extraction \cite{d3}.  
High data volume of point cloud representation
poses challenges to data transport and storage.  One   important point cloud processing task
involves resampling for reducing the data size
in a point cloud
while preserving vital 3D structural features. Point cloud resampling serves important practical
applications such as segmentation, object classification, and data compression.

The literature contains a variety of works
on point cloud resampling.
A centroidal Voronoi tessellation based method in \cite{c1} can progressively generate
high-quality resampling results with isotropic or 
anisotropic distributions from a given point cloud. 
The 
3D downsampling technique based on a growing neural gas network developed in \cite{c2} 
deals with data outliers and can
represent 3D spaces by an induced Delaunay 
triangulation of the input space. Importantly, graph-based resampling has exhibited attractive capability to capture 
some underlying structures of point clouds \cite{f1,z1}. 
For example, the work \cite{c3} focused on computationally efficient resampling and proposed several graph-based 
filters to capture the distribution of point data. 
Another graph-based example is 
a graph-based contour-enhanced resampling method 
presented in \cite{d4}. 
We also note a different, feature-based, approach to point cloud resampling based on edge detection and feature extraction. 
In \cite{c4}, the authors presented a sharp feature detector using Gaussian map clustering on local neighborhoods. 
Bazazian and co-authors \cite{c5} extended this principle by leveraging principal component analysis (PCA)
to develop a new agglomerative clustering method. This work establishes speed and accuracy improvement
by analyzing the eigenvalues of covariance matrix 
defined by $k-$nearest neighbors. 

Despite these and other successes,  graph-based and feature-based resampling methods still need to
overcome some limitations. 
For graph-based methods, each graph edge can only model pairwise relationships 
even though multilateral interactions
of point data represent highly informative 
characteristics of 3D objects. 
For example,  bilateral connections only
partially represent multilateral relationship among 
multiple points on the same surface. 
Another open issue relates to
the efficient construction of a graph 
for an arbitrary point cloud.
Among feature-based methods,  
how to select adequate features and 
filter parameters remains a practical design
challenge. The development of 
a general model with robust parameter selection 
remains an open problem.

There have been some reported successes at applying
hypergraphs to model and characterize the multilateral interactions 
among multimedia data points \cite{d5}. Hypergraph consists of nodes and hyperedges. Since each hyperedge can include more than two nodes,  hypergraph provides a more general 
tool in point cloud processing to model multilateral point relationships on object surfaces. Furthermore, as a
generalization of graph signal processing (GSP)  \cite{d9},
hypergraph signal processing (HGSP) \cite{d6,d8} has enjoyed
additional notable successes in 
processing point clouds for segmentation, sampling, and denoising \cite{d7,c11}.

In this paper, we propose a novel 3D point cloud 
resampling method using kernel filtering based on hypergraph signal processing. 
Leveraging hypergraph stationary process, we first estimate hypergraph kernel basis for a point cloud. 
We further define a local smoothness with respect to high-pass filtering in spectrum domain for 
resampling. 
In order to test the model preserving 
property on complex models, 
we develop a simple method for point cloud recovery based on alpha complex and Poisson sampling. 
Our experimental results demonstrate the compression efficiency and robustness of our proposed point-cloud resampling method.

\section{Background}

\subsection{Point Cloud}

A point cloud is a collection of point coordinates in 3D space. In point clouds, each point consists of 
three coordinates 
and may contain additional features, such as colors and normals \cite{c6}. In this work, we focus on the coordinates of these points when
resampling point clouds. These point coordinates are represented by the location matrix $\mathbf{p}=[\mathbf{p}_1^T \cdots \mathbf{p}_N^T]^T$, where $N$ is the number of points in the point cloud and $\mathbf{p}_i$ is the 3D coordinates of the $i$-th point.

\subsection{Hypergraph Signal Processing}

Hypergraph signal processing (HGSP) is a framework based on the tensor representation of hypergraphs to capture high-order signal interactions \cite{d6}. In this framework, an $M$th-order $N$-dimensional representing tensor $\mathbf{A}=(a_{i_1 i_2 \cdots i_M})\in \mathbb{R}^{N^M}$represents a hypergraph with $N$ vertices, where each hyperedge is capable of connecting at most $M$ nodes.
Weights of hyperedges connecting fewer than $M$ nodes 
are normalized according to combinations and permutations \cite{d11}.

Orthogonal CANDECOMP/PARAFAC (CP) method enables
approximate decomposition of a representing tensor into
$\mathbf{A}\approx\sum_{r=1}^N \lambda_r \cdot \underbrace{\mathbf{f}_r \circ \ldots \circ \mathbf{f}_r}_{M \text{ times}}$,
where $\circ$ denotes tensor outer product, $\{\mathbf{f}_1,\cdots,\mathbf{f}_N\}$ are orthonormal basis to represent spectrum components, and $\lambda_r$ is corresponding frequency coefficient. Spectrum components \{$\mathbf{f}_1,\cdots,\mathbf{f}_N$\} form the full hypergraph spectral space. Other HGSP concepts, such as filtering, hypergraph Fourier transform, and sampling theory can be defined \cite{d6}.

\section{HGSP Point Cloud Resampling}

In this section, we introduce our kernel-based resampling method. Generally, our proposed method consists of three main steps: i) hypergraph spectrum construction, ii) 
Fourier transform, iii) high-pass filtering and resampling.

We first estimate hypergraph spectrum based on hypergraph stationary process. Shown in \cite{c11}, a stochastic signal $\mathbf{x}$ is weak-sense stationary 
iff it has zero-mean and its covariance matrix has the same eigenvectors as the hypergraph spectrum basis, i.e., $\mathbb{E}[\mathbf{x}] =\mathbf{0}$ and $\mathbb{E}[\mathbf{x}\mathbf{x}^H]=\mathbf{V}\Sigma_\mathbf{x}\mathbf{V}^{H}, \label{s2}$ where $\mathbf{V}$ is the hypergraph spectrum. We can estimate the hypergraph spectrum from the covariance matrix of three coordinates by assuming the signals to be stationary. Here, we use the same spectrum estimation strategy as \cite{c11}, and consider the adjacency tensor as a third-order tensor which is the minimal number of nodes to form a surface.

We define a $k\times k \times k$ 3D cube as the kernel to define a local signal $\mathbf{s}_i$. For simplicity, we use a $3\times 3 \times 3$ cube as the kernel in this paper, although larger kernel  
can facilitate flexibility in filter design 
at the cost of
complexity. Let $d$ be the center distance between 
two nearby voxels in the kernel. A proper selection of $d$ should allow $\mathbf{s}_i$ to capture local geometric information. If $d$ is too small, only a few neighbors of  the $i$-th point are included 
in the $\mathbf{s}_i$, leading to noise-sensitivity. If $d$ is too large, many neighbors are included in each voxel, which may lead to blurred 
local geometric information. A typical choice of $d$ is the intrinsic resolution $d^{(i)}$ of a point cloud. {We place the kernel towards the bounding box of the point cloud in our experiment.}

Let $\mathbf{s}=[\mathbf{X_1}\ \mathbf{X_2}\ \mathbf{X_3}]\in \mathbb{R}^{N_k\times3}$ be the coordinates of the total $N_k$ voxel centers in the kernel. For our selected $3\times 3\times 3$ kernel, $N_k=27$. We then normalize the coordinates $\mathbf{s}$ to 
define a zero mean signal $\mathbf{s'}$. By calculating eigenvectors $\mathbf{f}_1,\cdots,\mathbf{f}_{N_k}$ for $R_{\mathbf{s'}}=\mathbf{s'}\mathbf{s'}^T$, we can estimate hypergraph spectrum $\mathbf{V}=[\mathbf{f}_1,\cdots,\mathbf{f}_{N_k}]$.

\begin{algorithm}[t]
	\caption{Hypergraph Kernel-based Resampling}
	\label{alg:A}
	\begin{algorithmic}
		\STATE {\textbf{Input}: A 3D point clouds with $N$ nodes characterized by its coordinates $\mathbf{p}=[\mathbf{p}_1^T \cdots \mathbf{p}_N^T]^T$ of the point cloud, resampling ratio $\alpha$}.
		\STATE {\textbf{1.}} {Calculate the intrinsic resolution $d^{(i)}$ of point cloud};
		\STATE {\textbf{2.}} {Use the intrinsic resolution $d^{(i)}$ as $d$ to get the coordinates $\mathbf{s}=[\mathbf{X_1}\ \mathbf{X_2}\ \mathbf{X_3}]$ of voxel centers in the kernel};
		\STATE {\textbf{3.}} {Calculate the mean $\mathbf{\bar{s}}$ of each row in $\mathbf{s}$};
		\STATE {\textbf{4.}} {Normalize the original point cloud data as zero-mean in each row, i.e., $\mathbf{s'}=[\mathbf{X_1-\bar{s}}\ \mathbf{X_2-\bar{s}}\ \mathbf{X_3-\bar{s}}]$};
		\STATE {\textbf{5.}} {Calculate the eigenvectors $\mathbf{f}_1,\cdots,\mathbf{f}_{N_k}$, the eigenvalues $0\leq \lambda_1\leq \lambda_2 \leq \ldots \leq \lambda_{N_k}$ and the threshold $\theta$ of eigenvalues for $R_{\mathbf{s'}}=\mathbf{s'}\mathbf{s'}^T$};
		\FOR{$i=1,2,\cdots, N$}
		\STATE {\textbf{6.}}{Use $\mathbf{V}=[\mathbf{f}_1,\cdots,\mathbf{f}_{N_k}]$ to find $\hat{\mathbf{s}}_i$ based on Eq. (\ref{FT})};
		\STATE {\textbf{7.}}{Calculate the local smoothness $\sigma(\mathbf{p}_i)$ in Eq. (\ref{local smoothness})};
		\ENDFOR
		\STATE {\textbf{8.}} {Sort the local smoothness $\sigma(\mathbf{p}_i)$ and select the bottom $N_r=\alpha N$ points as the resampled point cloud}.
	\end{algorithmic}
\end{algorithm}

Next, for $i$-th point in the point cloud, its corresponding local signal $\mathbf{s}_i\in \mathbb{R}^{N_k}$ is defined according to the number of points in the voxel of kernel centered at point $i$. By focusing on 3rd-order tensors
and signal energies in spectrum domain, we can utilize hypergraph spectrum to calculate a simplified-form of hypergraph Fourier transform $\hat{\mathbf{s}}_i$ as
\begin{align}\label{FT}
	{\hat{\mathbf{s}}_i=\mathbf{V}^T\mathbf{s}_i}.
\end{align}

For edge-preserving purpose, we would like to separate high frequency coefficients from low frequency coefficients in spectrum domain. Since the spectrum bases corresponding to smaller $\lambda$ represent higher frequency components\cite{d6}, we can sort eigenvalues of $R_{\mathbf{s'}}$ as $0\leq \lambda_1\leq \lambda_2 \leq \ldots \leq \lambda_{N_k}$ with corresponding eigenvectors \{$\mathbf{f}_1,\cdots,\mathbf{f}_{N_k}$\}. We can then devise a threshold $\theta$ for dividing high frequency components and low frequency components based
on a sharp change between two adjacent eigenvalues. 

Given a threshold selection of $\theta$, we could further define a local smoothness $\sigma(\mathbf{p}_i)$ to select resampled points:
\begin{align}\label{local smoothness}
	{\sigma(\mathbf{p}_i)=\frac{\sum_{k \in \{1,2,\cdots, \theta\}}\vert \hat{\mathbf{s}}_i(k)\vert}{\sum_{k \in \{1, \cdots, {N_k}\}}\vert \hat{\mathbf{s}}_i(k)\vert}},
\end{align}
which is the fraction of high frequency energy. 

Finally, we resample the point cloud by selecting points with smaller $\sigma(\mathbf{p}_i)$, which corresponds to points containing larger amount of high frequency in the hypergraph. We summarize our algorithm as Algorithm \ref{alg:A}. Since the resampled point clouds tend to favor high-frequency points, they often contain sharper features and are less smooth. {Although the higher frequency basis vectors do not necessarily display a clear regular pattern, as shown in Fig.~\ref{fig:a1}, they have larger total variation over the (hyper)graph, which are corresponding to sharper features}. {Interested readers could refer to \cite{d6,d9,c11} for a comprehensive discussion on the high frequency features in hypergraphs.}

{For an unorganized point cloud, the computational complexity of our method amounts to $O(N^2+N\log{N}+ N_k(N_k+1)N+N_k^3)$, whereas for an organized point cloud, the complexity order is $O(N\log{N}+(N_k^2+N_k+1)N+N_k^3)$.}

\begin{table*}[t]
	\centering
	\caption{Numerical Results of Methods for Edge Preserving Resampling without and with Gaussian Noise.}
	\begin{minipage}[t]{0.53\textwidth}
	\scriptsize
		\centering
		\begin{tabular}{|c||c|c|c||c|c|c|}
			\hline
			\multicolumn{7}{|c|}{Combination of cubes}\\
			\hline
			& \multicolumn{3}{|c||}{Kernel Filtering} & \multicolumn{3}{|c|}{Fast Resampling}\\
			\hline
			Noise Level & Precision &  Recall & F1-Score & Precision & Recall & F1-Score\\
			\hline
			No noise & 0.3545 & 0.9332 & 0.5125 & 0.3827 & 1 & 0.5522\\
			\hline
			20\% & 0.2556 & 0.6689 & 0.3690 & 0.2326 & 0.6086 & 0.3357\\
			\hline
			40\% & 0.1052 & 0.2745 & 0.1518 & 0.0995 & 0.2600 & 0.1436\\
			\hline	
		\end{tabular}
	\end{minipage}
	\hspace{-7.2mm}
	\begin{minipage}[t]{0.46\textwidth}
	\scriptsize
		\centering
		\begin{tabular}{|c|c|c||c|c|c|}
			\hline
			\multicolumn{6}{|c|}{Cylinder}\\
			\hline
			\multicolumn{3}{|c||}{Kernel Filtering} & \multicolumn{3}{|c|}{Fast Resampling}\\
			\hline
			Precision & Recall & F1-Score & Precision & Recall & F1-Score\\
			\hline
			0.2212 & 0.9298 & 0.3524 & 0.2523 & 1 & 0.3981\\
			\hline
			0.1667 & 0.6834 & 0.2647 & 0.1497 & 0.5912 & 0.2359\\
			\hline
			0.0894 & 0.3566 & 0.1412 & 0.0064 & 0.2515 & 0.1003\\
			\hline		
		\end{tabular}
	\end{minipage}
	
	\begin{minipage}[t]{1\linewidth}
	\centering
	\scriptsize
		\captionsetup{labelformat=parens}
		\begin{tabular}{|c||c|c|c||c|c|c|}
			\hline
			\multicolumn{7}{|c|}{Pyramid}\\
			\hline
			& \multicolumn{3}{|c||}{Kernel Filtering} & \multicolumn{3}{|c|}{Fast Resampling}\\
			\hline
			Noise Level & Precision & Recall & F1-Score & Precision & Recall & F1-Score\\
			\hline
			No noise & 0.4567 & 0.6860 & 0.5097 & 0.6750 & 0.9034 & 0.7168\\
			\hline
			20\% & 0.4191 & 0.5871 & 0.4538 & 0.4300 & 0.5696 & 0.4537\\
			\hline
			40\% & 0.3017 & 0.4024 & 0.3205 & 0.2345 & 0.2874 & 0.2395\\
			\hline
		\end{tabular}
	\end{minipage}
	\label{table:1}
\end{table*}

\section{Experimental results}

We now describe our experiment setup and present test results of the proposed resampling algorithm.

\subsection{Edge Preserving on Simple Synthetic Point Clouds}

\begin{figure}
	\centering
	\subfloat{
		\includegraphics[width=.12\textwidth]{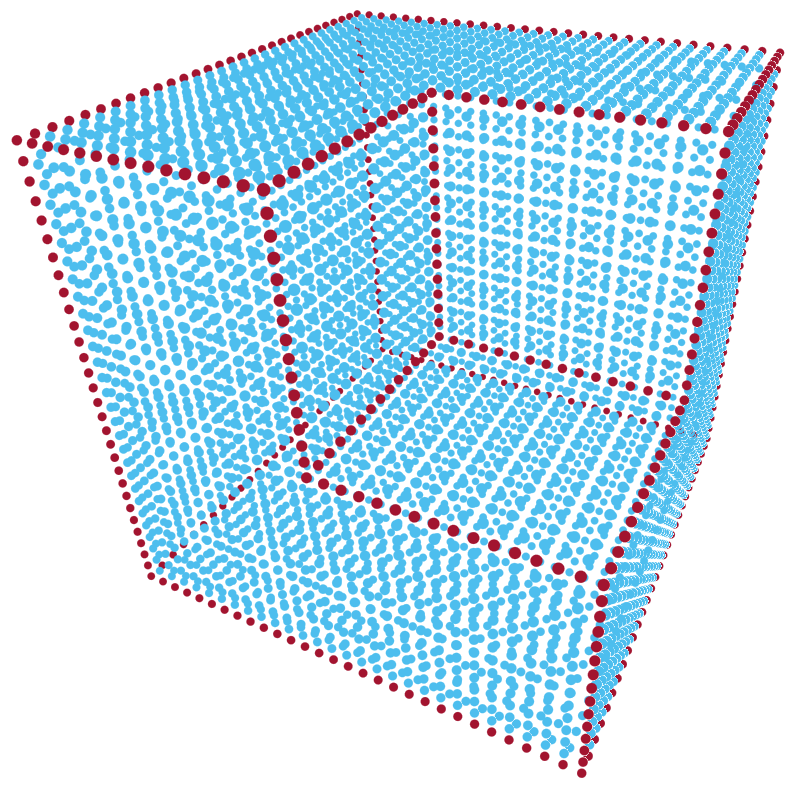}
		\label{Fig:11}
	}
	\subfloat{
		\includegraphics[width=.12\textwidth]{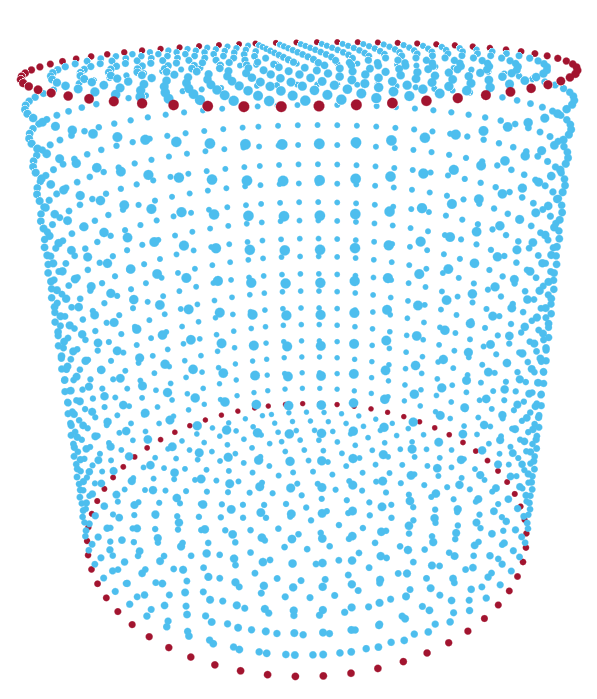}
		\label{Fig:12}
	}
	\subfloat{
		\includegraphics[width=.12\textwidth]{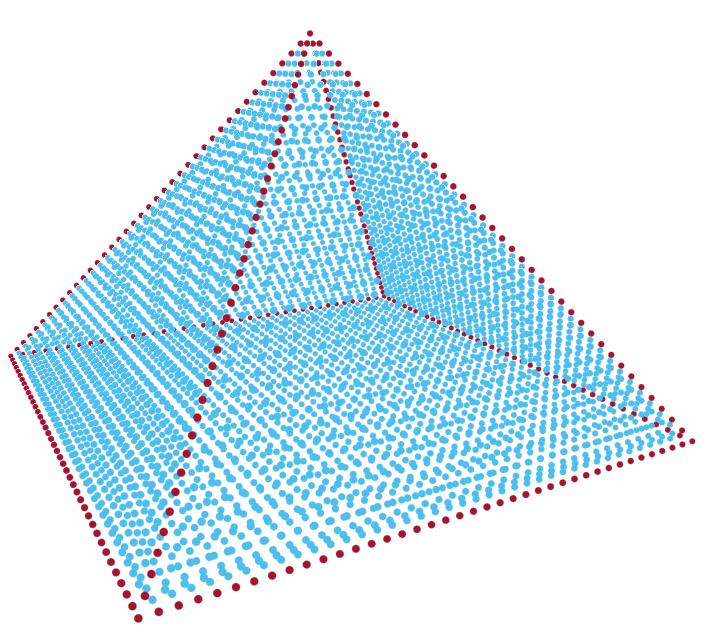}
		\label{Fig:13}
	}
	\caption{Examples of synthetic point clouds with labeled edge.}
	\label{Fig:1}
\end{figure}

\begin{figure}
    \centering
    \includegraphics[width=2in]{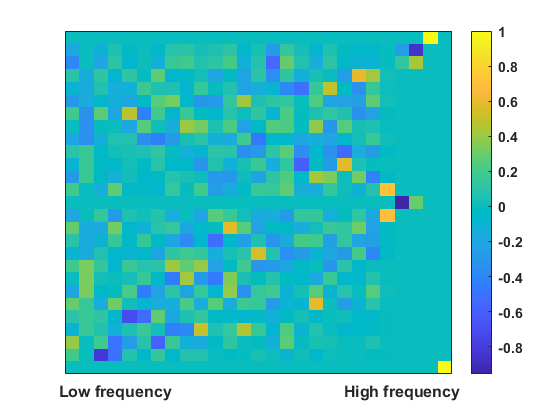}
    \caption{An example of graph spectrum matrix of the kernel. Each column of the matrix corresponds to one frequency.}
    \label{fig:a1}
\end{figure}

As we mentioned in Section \uppercase\expandafter{\romannumeral3}, one important resampling objective is to preserve sharp features in a point cloud such as edges and corners. Here, we study the edge preserving capability of our proposed algorithms by testing over several simple synthetic point clouds. The benefit for selecting synthetic point clouds is
the ability to take advantage of known ground truth regarding edges and our ability to label them. We generate these synthetic point clouds by uniformly 
sampling external surface of models constructed from cylinders, pyramids, and combinations of cubes 
of various sizes. Examples of synthetic point clouds are shown in {Fig. \ref{Fig:1}}, where the points on edges are marked with red and the rest points are in blue.

To measure the accuracy of the edge preserving features, we use the $F_1$ score defined as:
\begin{align}\label{F1}
	F_1 = 2\times\frac{\rm{Precision} \times \rm{Recall}}{\rm{Precision} + \rm{Recall}},
\end{align}
in which precision is 
the proportion of points correctly preserved by the resampling algorithm whereas the recall rate is the 
ratio of correctly preserved edge points to
all ground truth edge points.

We compare our method with a graph-based fast resampling method given in \cite{c3}. We set resampling ratio to $\alpha=0.2$ for all point clouds. More results on different sampling ratios are in Section \uppercase\expandafter{\romannumeral4}-C. The parameters of fast resampling method are set to typical values 
according to \cite{c3}. In order to study  robustness, we include random Gaussian noise of $\mathcal{N}(0,(0.2d^{(i)})^2)$ and $\mathcal{N}(0,(0.4d^{(i)})^2)$ to the coordinates of point cloud. The
noise effects on the
results are shown in {Table \ref{table:1}}. The test results show that our hypergraph kernel-based filtering method achieves better performance in preserving edges in noisy data when compared against GSP-based method. {Since the realistic point clouds always contain noise, our proposed method is more practical in real applications.}

\subsection{Edge Detection on Complex Real Point Cloud}

\begin{figure}[t]
	\centering
	\subfloat{
		\centering
		\includegraphics[width=.45\textwidth]{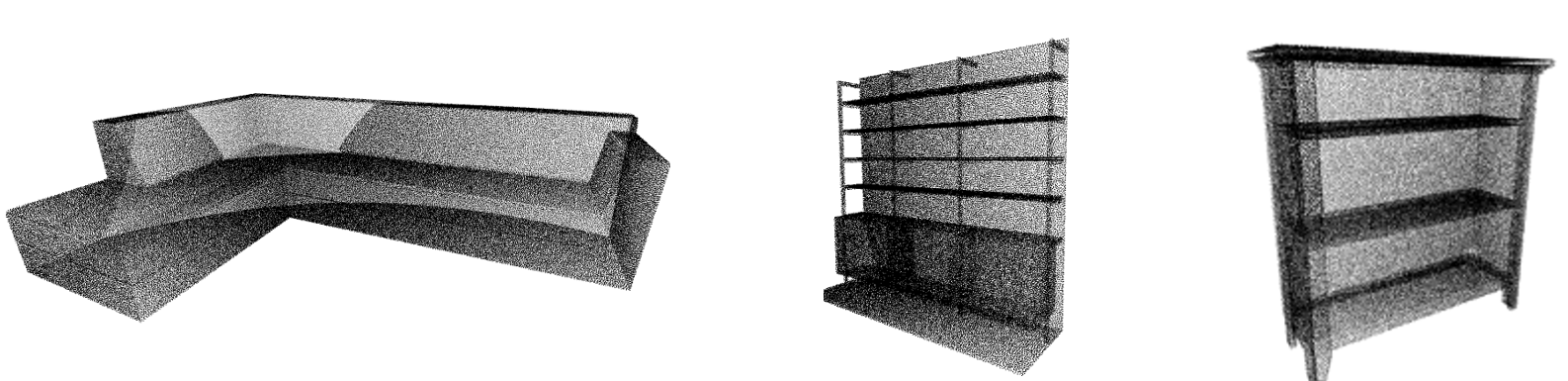}
		\label{fig1}
	}
	\\
	\subfloat{
		\centering
		\includegraphics[width=.45\textwidth]{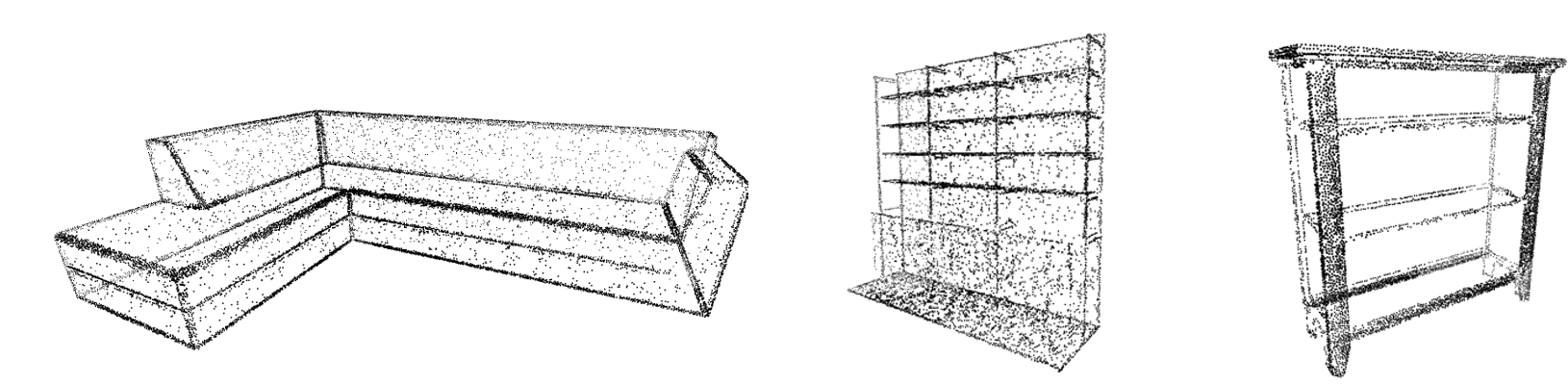}
		\label{fig2}
	}
	\caption{Examples of edge detection for real point cloud.}
	\label{fig11}
\end{figure}

To test our proposed algorithm in a more general setting, we also implement edge-detection based on our resampled data in more complex practical point clouds. For these datasets, there are no known ground truth edges. Therefore, we shall present the matching point clouds in Fig. \ref{fig11}, where the top row contains 
the 3 original point clouds and the bottom row
shows the resampled point clouds. 
Visual inspection shows that our resampling method 
easily captures contours of the objects.

\begin{table*}[t]
	\centering
	\captionsetup{justification=centering, labelsep=newline,font=sc}
	\caption{Mean distance between the best point cloud and the corresponding original point cloud with $\alpha=0.2$.}
	\vspace{-2mm}
	\begin{tabular}{|c||c|c|c|c|}
		\hline
		Categories & Kernel Filtering & Fast Resampling & Eigenvalues Analysis & PCA and Clustering\\
		\hline
		Cap & 0.0101 & 0.0102 & \textbf{0.0087} & 0.0115 \\
		\hline
		Chair & \textbf{0.0113} & 0.0118 & 0.0125 & 0.0126\\
		\hline
		Mug & \textbf{0.0134} & 0.0141 & 0.0150 & 0.0147\\
		\hline
		Rocket & \textbf{0.0069} & 0.0070 & 0.0070 & 0.0073\\
		\hline
		Skateboard & 0.0080 & \textbf{0.0079} & 0.0082 & 0.0083\\
		\hline
	\end{tabular}
	\label{table:2}
\end{table*}

\subsection{Point Cloud Recovery from Resampled Point Clouds}

To study the model preservation of our proposed method, we recover original point clouds after resampling 
and evaluate the difference between original and the 
recovered point clouds.

\textbf{Recovery Method:} The point cloud recovery contains two steps: 1) reconstruct the surface of object from  resampled point cloud, and 2) sample on reconstructed surface to generate the recovered point cloud. Since many points in the edge preserving resampled point clouds are located in areas with high local variation, such as surface edges, the resampled point clouds is non-uniform, as shown in {Fig.~\ref{fig:downsamplept}}. Therefore, most commonly used surface reconstruction methods such as Poisson reconstruction \cite{e1} may not be compatible with
these point clouds. In order to reconstruct surfaces, we build alpha complex \cite{e2} of the resampled point cloud. To minimize the sensitivity to reconstruction, 
we reconstruct six different surface models based on different parameters for each resampled point cloud. 
We then apply Poisson-disk resampling to sample over the alpha complex to recover the original point cloud. Only the recovered point cloud with the smallest distance to the original point cloud is kept as output.

\textbf{Distance between two point clouds:} To measure the similarity between the original and recovered point clouds, we define a distance metric
\begin{align}\label{distance}
	D_0=\frac{1}{N_1}\Sigma_{i=1}^{N_1} 
	\min_{j: \|\mathbf{p}_i-\mathbf{p}_{c,j}\|<d_\theta} \|\mathbf{p}_i-\mathbf{p}_{c,j}\|,
\end{align}
where $N_1$ is the number of points in the original point cloud, $\mathbf{p}_i$ and $\mathbf{p}_{c,j}$ are 
points on the original and recovered point clouds, respectively. Although we apply different parameters in the reconstruction, errors such as phantom surfaces that 
do not exist in the original point cloud may still
appear. Therefore, we select a threshold $d_\theta$ to
three times the intrinsic resolution of the original point cloud. $D_0$ is the average distance among points 
of the original point cloud within $d_\theta$ from the closest points in the best recovered point cloud.

\begin{figure}
	\centering
	\includegraphics[width=2.5in]{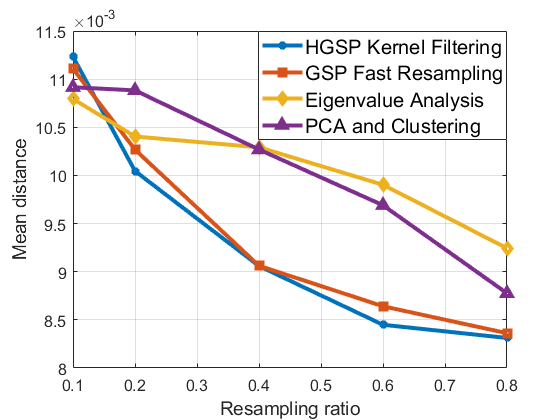}
	\caption{Performance of Recovery with Different Resampling Ratio.}
	\label{fig:Curve}
\end{figure}

\begin{figure}[t]
	\centering
	\subfloat[Original Cap]{
		\centering
		\includegraphics[width=.14\textwidth]{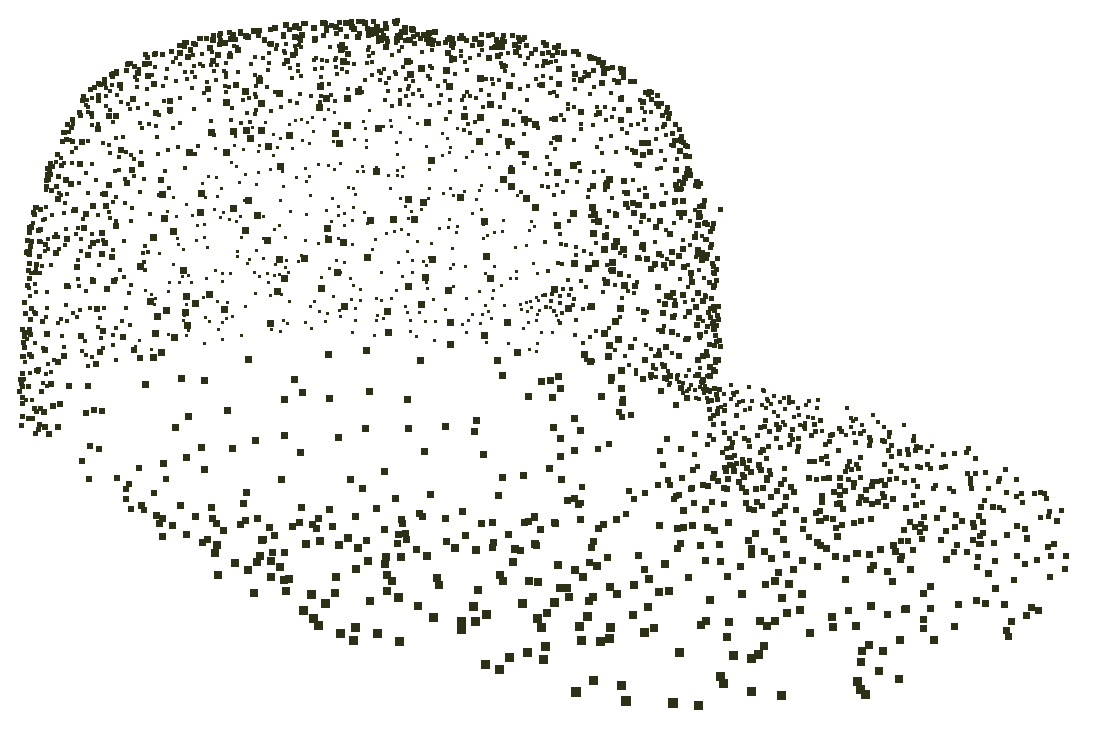}
		\label{fig:originalpt}
	}
	\hfill
	\subfloat[Resampled]{
		\centering
		\includegraphics[width=.14\textwidth]{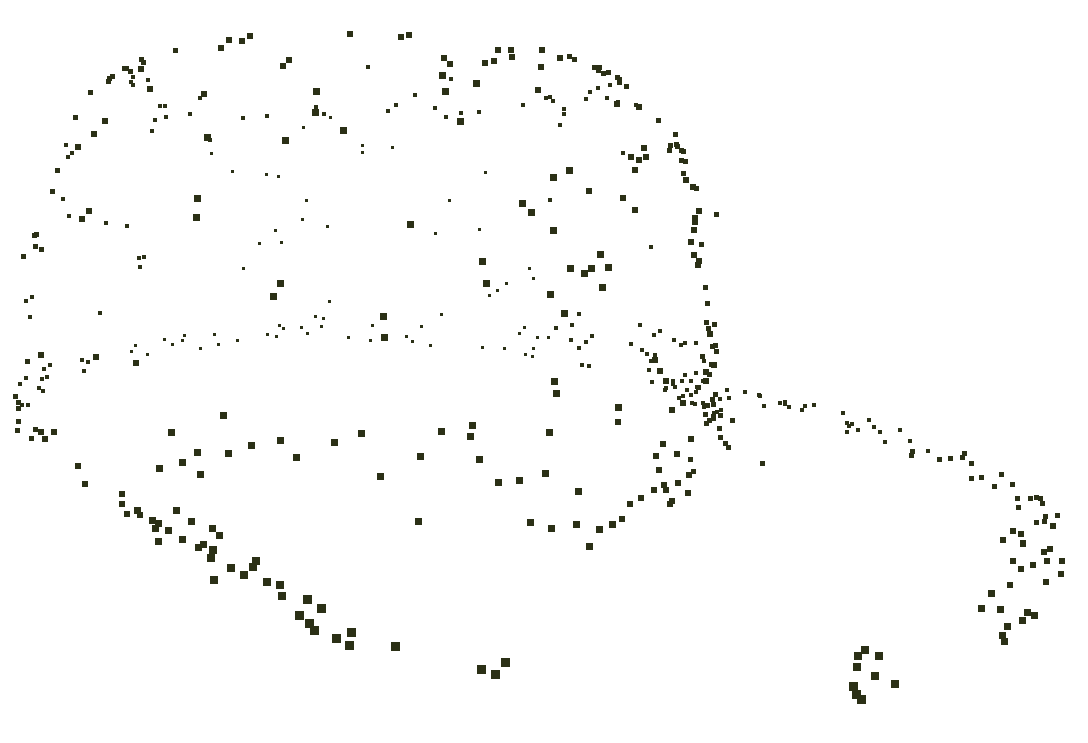}
		\label{fig:downsamplept}
	}
	\hfill
	\subfloat[Recovered]{
		\centering
		\includegraphics[width=.14\textwidth]{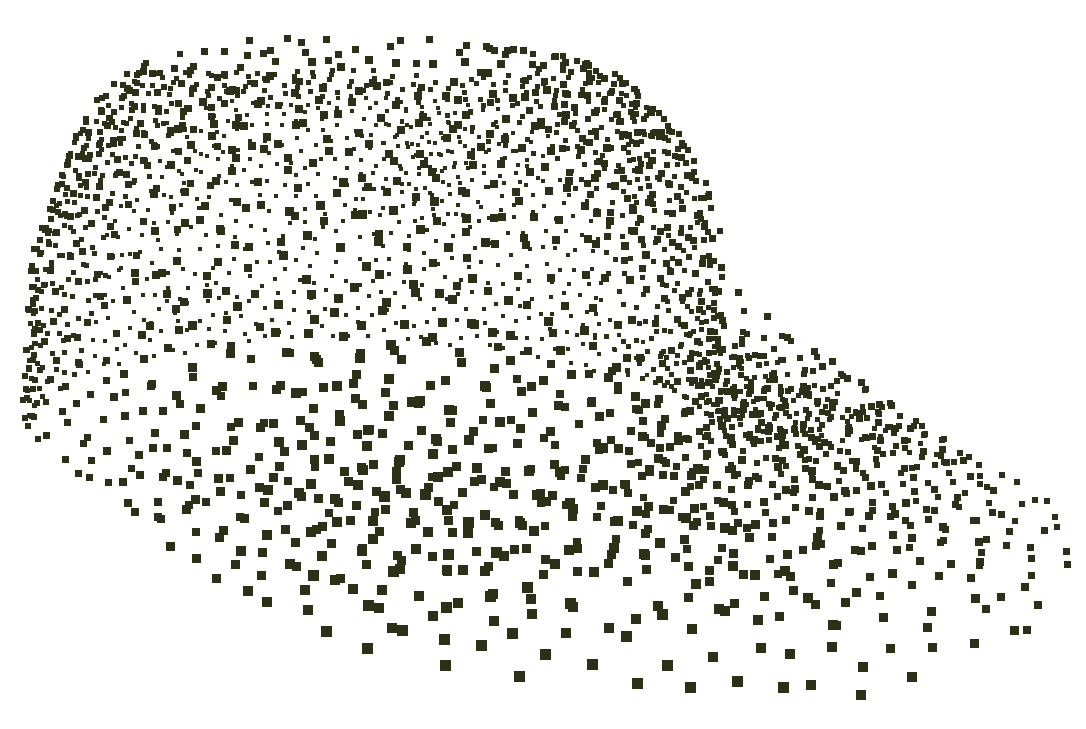}
		\label{fig:recoverpt}
	}
	
	\caption{Example of Recovered Point Cloud.}
	\vspace{-4mm}
\end{figure}
\textbf{Visual and numerical results.} 

\begin{figure}[t]
	\centering
	\subfloat[Original]{
		\centering
		\includegraphics[width=.14\textwidth]{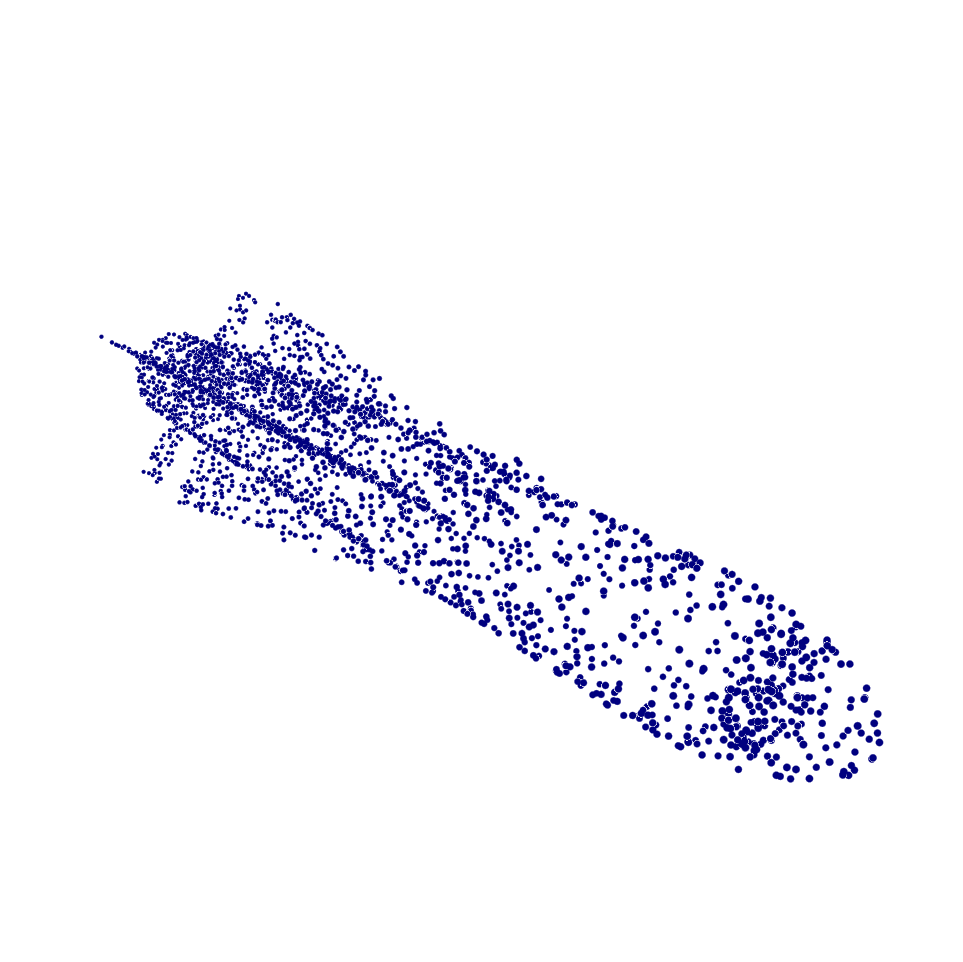}
		\label{fig:5a}
	}
	\hfill
	\subfloat[Our resampled result]{
		\centering
		\includegraphics[width=.14\textwidth]{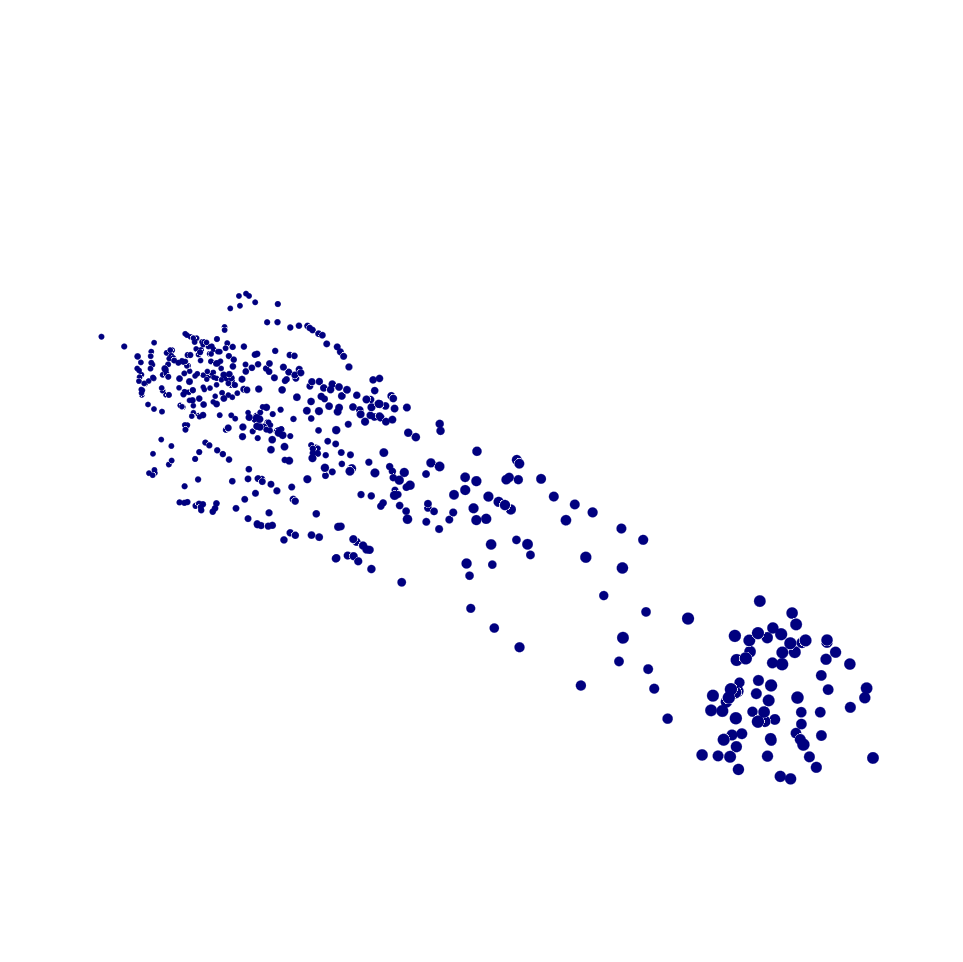}
		\label{fig:5b}
	}
	\hfill
	\subfloat[GSP resampling]{
		\centering
		\includegraphics[width=.14\textwidth]{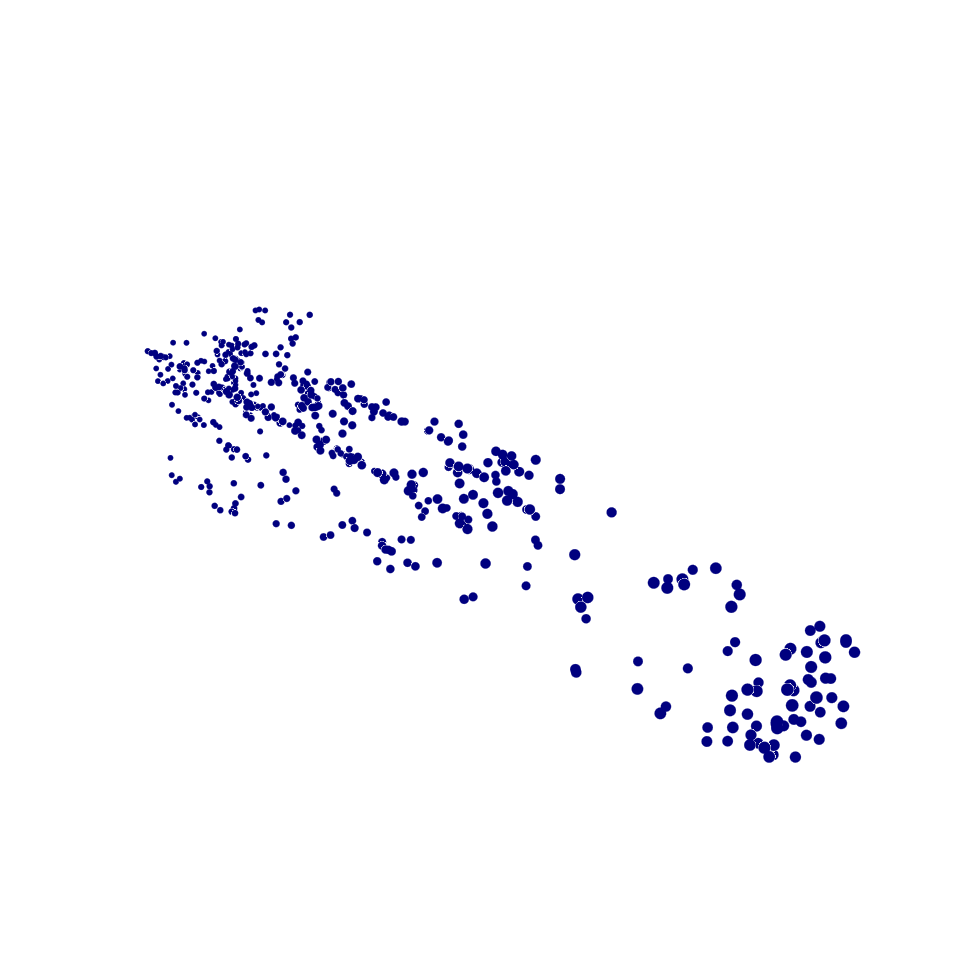}
		\label{fig:5c}
	}
	\vspace{-2mm}
	\caption{Example of resampled results comparison.}
	\label{fig:5}
\end{figure}

Our experiment uses models in six different categories in Shapenet 
\cite{c12} as the original point clouds. We compare our hypergraph kernel-based filtering method with the GSP-based method in \cite{c3}. We also compare to two edge detection methods, i.e., eigenvalues analysis and PCA clustering methods in \cite{c5}. We resample the same number of nodes for all methods under comparison.

Our experiment follows the following steps: {f}irst, we use resampling and edge detection methods to calculate the resampled point clouds at
different values of
resampling ratio $\alpha$; {n}ext, we use the proposed recovery method to generate the recovered point clouds; {w}e then calculate the distance between the recovered point clouds and the corresponding original ones; {f}inally, we calculate the mean distance between the best recovered point cloud and their original point cloud for each method. An example of the recovered point cloud is shown in {Fig.~\ref{fig:recoverpt}}. Numerical results for $\alpha=0.2$ are shown in {Table~\ref{table:2}}. Our test results
show that the
proposed method achieves lower mean distance 
than that from GSP-based resampling. 
When compared with edge detection methods, 
our proposed algorithm 
generates similar or smaller mean distance in most categories. {Fig.~\ref{fig:5} shows another example of point cloud resampling. The edges of the rocket tail are well preserved in our resampled results as shown in Fig.~5(b), while the GSP-based method fails to show those tails.} We also examine the effect of different resampling 
ratio $\alpha$ in Fig.~\ref{fig:Curve} 
by using the same set of 
point clouds. Our proposed method exhibits
superior performance
over traditional methods in terms of mean distance for 
various resampling ratios. {In terms of computation, we process the point cloud with 349,300 points in 50.88 seconds by using Matlab on a standard desktop (4.4 GHz Intel Core i5, 64 GB memory) without parallel processing, while the GSP-based method needs 56.82 seconds.} These tests demonstrate the efficiency and efficacy
of our point cloud recovery. It also shows that 
HGSP can model the point clouds more efficiently
than regular GSP in some point cloud applications. 

\vspace{-2mm}
\section{Conclusion}

In this work,
we developed a 3D point cloud resampling method 
using kernel filtering based on hypergraph 
signal processing (HGSP). Our
proposed HGSP resampling is simple and 
effective. Experimental results demonstrated that our 
HGSP method outperforms traditional 
graph-based methods or feature-based methods 
such as edge detection in 
terms of robustness to the noise and 
model preservation.  
This work establishes HGSP as an efficient tool to model multilateral relationship and to extract features
in point cloud applications. 

\bibliographystyle{IEEEbib}

\end{document}